\documentclass[journal=acsnano,manuscript=article]{achemso}

\usepackage[version=3]{mhchem} 
\usepackage{graphicx}
\usepackage{amsmath,amssymb}
\usepackage{upgreek}
\usepackage{color}
\usepackage{xcolor}
\usepackage{miller}           
\usepackage{float}
\usepackage{multirow}
\usepackage{rotating}
\usepackage{textcomp}
\usepackage{units}
\usepackage{comment}
\usepackage[english]{babel}
\usepackage{hyperref}

\hypersetup{
    colorlinks=true,
    linkcolor=black,
    citecolor=blue,
    urlcolor=blue
}

\newcommand{\degC}{\ensuremath{^{\circ}\mathrm{C}}}

\title{From stripes to hexagons: strain-induced 2D Pb phases confined between graphene and SiC}

\author{Markus Gruschwitz}
\affiliation{Solid Surface Analysis, Institute of Physics, Chemnitz University of Technology, 09126 Chemnitz, Germany}

\author{Sergii Sologub}
\affiliation{Solid Surface Analysis, Institute of Physics, Chemnitz University of Technology, 09126 Chemnitz, Germany}
\alsoaffiliation{Institute of Physics, National Academy of Sciences of Ukraine, Nauki Avenue 46, 03028 Kyiv, Ukraine}

\author{Chitran Ghosal}
\affiliation{Solid Surface Analysis, Institute of Physics, Chemnitz University of Technology, 09126 Chemnitz, Germany}

\author{Zamin Mamiyev}
\affiliation{Solid Surface Analysis, Institute of Physics, Chemnitz University of Technology, 09126 Chemnitz, Germany}

\author{Yuran Niu}
\affiliation{MAX IV Laboratory, Lund University, Fotongatan 2, 22484 Lund, Sweden}

\author{Alexei Zakharov}
\affiliation{MAX IV Laboratory, Lund University, Fotongatan 2, 22484 Lund, Sweden}

\author{Christoph Tegenkamp}
\email{christoph.tegenkamp@physik.tu-chemnitz.de}
\affiliation{Solid Surface Analysis, Institute of Physics, Chemnitz University of Technology, 09126 Chemnitz, Germany}

\date{\today}

\begin{document}

\begin{abstract}
The intercalation of metals beneath graphene offers a powerful route to stabilizing and protecting novel two-dimensional (2D) phases. The epitaxial growth of Pb monolayers on SiC(0001), combined with the relatively large spacing of the suspended graphene, makes this system particularly distinctive. Using low-energy electron diffraction (LEED) and various microscopy techniques—including scanning electron microscopy (SEM), scanning tunneling microscopy (STM), and low-energy electron microscopy (LEEM)—we have investigated the intercalation process across multiple length scales. Our analysis reveals the formation of different 2D Pb monolayer phases, such as stripes and hexagons, which emerge due to the interplay between substrate pinning and strain within the Pb layer, depending on local coverage. These findings provide new insights into the strain-driven stabilization of intercalated metal layers and highlight the potential of graphene as a versatile platform for engineering low-dimensional materials.
\end{abstract}

\section{Introduction}

Two-dimensional (2D) material systems offer a unique platform for exploring emergent quantum phenomena at the atomic scale. Among these, the intercalation of metal atoms beneath graphene has proven to be a powerful approach to engineering new structural and electronic phases \cite{Berger2020,Sutter2010,Forti2011}. Intercalants can decouple the graphene from its substrate, modify charge transfer, or even induce magnetism and superconductivity, depending on their nature and the interfacial configuration \cite{Dedkov2010,Yao2014}. 

The case of lead (Pb) is particularly interesting due to its rich phase diagram and the emergence of superconductivity in atomically thin 2D Pb layers on Si surfaces, robust quantum size effects in thin films, and its ability to induce strong spin–orbit coupling in graphene \cite{Zhang1998,Luh2001,Brun2009}.
Pb overlayers on semiconductor substrates such as Si(111) have long served as a model system for studying confined electronic states, where discrete quantum well states emerge due to electron confinement along the growth direction. These effects, commonly referred to as ``electronic growth'' \cite{Zhang1998,Yakes2004,Hupalo2003,Jeffrey2007}, arise from the delicate balance between surface energy, quantum confinement, and strain. In ultra-thin Pb films, such confinement leads to quantized energy levels and strongly modulates the film’s stability, morphology, and superconducting properties \cite{Brun2009}. When confined beneath a 2D layer like graphene, additional effects come into play: graphene not only acts as a capping layer protecting the intercalated phase from ambient conditions, but also modifies the boundary conditions for electron confinement and imposes additional constraints on the structural order and strain state of the metal layer \cite{Pacile2013,Forti2020}. Moreover, densely packed Pb monolayers on Si(111) exhibit pronounced spin-split surface states, which, on vicinal Si(557), undergo a topological transition forming a correlated spin–order state \cite{Brand2017, Quentin2020, Brand2015}.

In this context, the intercalation of Pb beneath epitaxial graphene on SiC(0001) presents a particularly rich system. The so-called zero-layer graphene (ZLG), a partially covalently bound carbon buffer layer, provides a defect-rich and reactive template for initiating intercalation. The relatively large spacing between suspended graphene and the SiC substrate allows for thicker and structurally relaxed intercalated layers, potentially accommodating multiple structural motifs \cite{Schaedlich2023}. Yet, the diffusion pathways, kinetics, and energy barriers involved in the intercalation process remain highly dependent on local defects, cluster size, and thermal treatment history \cite{Han2024,Wang2025}.

Furthermore, the confined Pb layer exhibits signatures of quantum confinement and electronic reconstruction, analogous to but distinct from those seen in free Pb monolayers on semiconductors \cite{Brun2009,Luh2001}. The intercalated Pb structures preserve aspects of the $\alpha$-Pb phase found on Si(111) while, in addition, the potential of the graphene overlayer is superimposed \cite{Schaedlich2023}.  These reconstructions likely modulate the local density of states and could enable future exploration of quantum phases in encapsulated environments. The combination of structural tunability and confinement makes intercalated Pb beneath graphene a compelling model system for studying strain-driven self-organization and quantum electronic phenomena in 2D metal layers.

In this work, we present a systematic intercalation study of Pb beneath ZLG on SiC(0001), combining scanning electron microscopy (SEM), low-energy electron microscopy (LEEM), scanning tunneling microscopy (STM), and spot-profile analysis low-energy electron diffraction (SPA-LEED). Our multiscale approach allows us to follow the evolution of Pb from initial cluster formation to its transformation into a confined, 2D interfacial layer. We observe distinct interfacial Pb structures—such as stripe-like domains and hexagonal phases (also known as bubbles \cite{Schaedlich2023})—whose formation is attributed to a subtle interplay between compressive and tensile strain within the Pb layer. These strain effects are closely linked to the intercalation protocol and effective Pb coverage.

\section{A. Results and Discussion: the process of intercalation}
\subsection{Quality of the 2D Pb interface layer studied by SPA-LEED}

\begin{figure}[tb]
	\begin{center}
		\includegraphics[width=\linewidth]{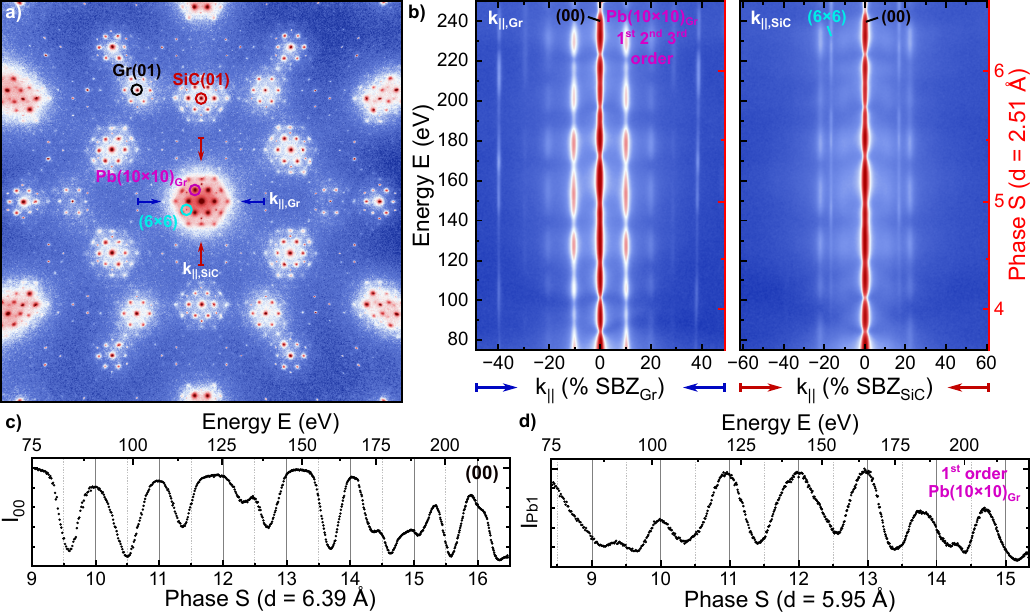}
		\caption{(a) Extended SPA-LEED pattern of an intercalated Pb monolayer showing the quasi-$(10\times10)_\text{Gr}$ reconstruction spots with respect to the graphene lattice ($E=225$~eV, $T\approx180$~K). (b) Reciprocal intensity map along the graphene (left, blue arrows in a) and SiC (right, red arrows in a) direction for different electron energies. The Pb-induced spots up to the third order are marked. The phase axis is calibrated for single SiC steps. (SBZ: surface Brillouin zone). The intensity changes of the (00) spot and the first-order Pb-induced superstructure spots are plotted in (c) and (d), respectively ($T=70$~K). The phase $S$ is derived with respect to the effective layer thickness of the graphene and Pb heterostructure \cite{Schaedlich2023}.
        \label{FIG1} 
		}
	\end{center}
\end{figure}

Figure \ref{FIG1}a) shows a typical SPA-LEED pattern after successful intercalation of Pb. In addition to the characteristic integer-order diffraction spots from the SiC(0001) substrate and the quasi-free-standing monolayer graphene (QFMLG), the diffraction pattern exhibits, besides the characteristic $(6\times6)$ spots  replica spots from substrate and graphene lattices, indicative for the long-range ordered phases. 
Indicators for incipient intercalation are the suppression of ZLG-related superstructure spots, i.e. $\left(6\sqrt{3}\times 6\sqrt{3}\right)R30^\circ$, increase of graphene spot intensity and appearance of the so-called bell-shaped background as it is closely related to the formation of extended 2D materials \cite{Chen2020} and is considered a hallmark of the decoupling of ZLG by intercalation \cite{Mamiyev2022}. Superstructure spots arising from Pb structures at the interface are observed near the $(10\times10)_\text{Gr}$ position relative to the graphene lattice \cite{Schaedlich2023}. The absence of Pb(111) diffraction spots rules out a simple moiré pattern between graphene and Pb as the origin of this Pb-induced superstructure.

\begin{figure}[tb]
\begin{center}
	\includegraphics[width=.8\linewidth]{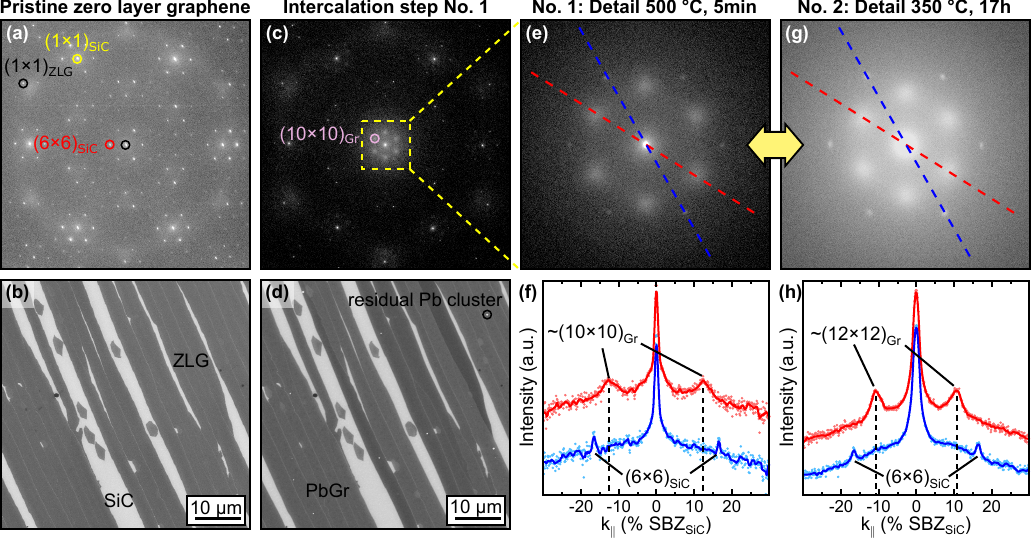}
	\caption{(a,b) SPA-LEED pattern ($E=168$~eV) and SEM (15~kV, 1~nA) image characterizing the center of the sample  before Pb intercalation.  (c) SPA-LEED pattern after the first intercalation cycle at medium temperatures (No.\ 1 in Tab.\ \ref{TAB1}). (d) SEM at the same area after the final preparation step (No.\ 6 in Tab. \ref{TAB1}). The central part of the SBZ$_\text{SiC}$  (yellow box, approximately 30\%) was separately measured after each intercalation step. (e,f) Detailed scan of marked box in (c) and linear intensity profiles intersecting $(6\times6)$ (blue) or superstructure spots (red). (g,h) Same measurements were repeated after the next intercalation step, including mild long-time annealing (No.\ 2 in Tab.~\ref{TAB1}). The superstructure spots increase significantly in intensity and shift to larger periodicity. All measurements were performed at 300~K. 
    \label{FIG2} 
   }
\end{center}
\end{figure}

To further characterize the intercalated Pb monolayer structure as well as the overlying QFMLG, we performed a profile analysis over a broad energy range. The intensity variations are shown exemplarily for one sample in panel (b). In total, three samples were investigated, which were found to be nearly identical with respect to the following observations: in addition to the (00)-spot, the first, second, and even third-order diffraction peaks of the Pb reconstruction are visible (left part of panel n). Most notably, the diffraction peaks of the intercalated phase exhibit pronounced intensity oscillations. To investigate this in more detail, the line profiles were analyzed.  Overall, the full width at half maximum (FWHM) values are very small (1~\% SBZ), and their variation with the scattering phase is weakly pronounced both indicating a well-ordered phase with long-range order extending across the transfer width of the SPA-LEED system (200~nm).
Panels (c) and (d) show the integrated intensities of the (00)- and first-order Pb superstructure spot, respectively. The best agreement for both spots is found for a layer spacing of 5.95--6.39~\AA. Interestingly, this oscillation correlates nicely with the effective layer spacing of the 2D heterostructure composed of the Pb monolayer and the graphene, as we recently demonstrated using XSW and STEM \cite{Schaedlich2023,Gruschwitz2024}. The pronounced intensity variations can thus be interpreted as a resonance effect governed by the heterostructure itself. The partial deviations from this fundamental oscillation are visible for both spots, particularly in the high-energy range, and are very likely due to multiple scattering effects. Nevertheless, the robust intensity oscillations clearly highlight the high structural quality and electronic homogeneity of the intercalated 2D Pb phase in agreement with the homogeneous contrast seen in the SEM shown below.

\subsection{Reversible phase transitions of the intercalated 2D Pb phase}

\begin{figure}[tb]
	\begin{center}
		\includegraphics[width=.5\columnwidth]{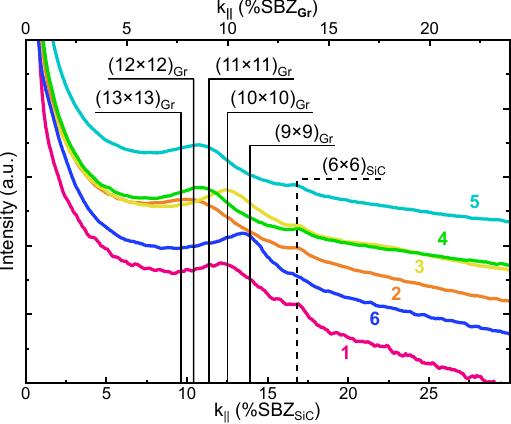}
		\caption{Angle-averaged radial intensity profiles of the diffraction patterns (central part of the SBZ$_\text{SiC}$, $E = 168$~eV, $T=300$~K) measured after successive intercalation cycles. The numbers refer to the steps explained in Tab.~\ref{TAB1}. Superstructure periodicities with respect to graphene and SiC are marked by solid and dashed lines, respectively.}
        \label{FIG3}
	\end{center}
\end{figure}

In order to achieve a deeper insight into the intercalation process, we monitored each step during a typical intercalation sequence of a ZLG sample in detail by high-resolution diffraction. A combination of deposition and annealing steps were performed, including varying coverages and temperatures,  in order to achieve homogeneous intercalation of the ZLG. The parameters of the intercalation sequence are listed in Tab.~\ref{TAB1}. 

Previous experiments revealed that the intercalation of Pb crucially depends on imperfections within the ZLG \cite{Gruschwitz2021, Ghosal2022, Vera2024} since direct penetration through the ZLG is energetically unfavorable \cite{Han2023}. Therefore, we deliberately chose an undergrown ZLG on the 6H-SiC(0001) sample for an accelerated intercalation. The initial  state of the ZLG sample examined by SPA-LEED and SEM is shown in Fig.~\ref{FIG2}(a,b), respectively. In addition to the characteristic reconstruction spots of ZLG, diffuse intensities at $\left(\sqrt{3}\times\sqrt{3}\right)_\text{SiC}$ positions appear consistent with a defective ZLG layer or Si-rich reconstructions at uncovered SiC terraces.  
The undergrown phase of the ZLG is also evident from the SEM contrast of light (SiC) and dark (ZLG) areas in Fig.~\ref{FIG2}(b). In fact, samples with undergrown ZLG underwent a comparatively fast Pb intercalation. The significant change of the diffraction pattern in Fig.~\ref{FIG2}(c) after the first intercalation cycle at medium temperatures (No.\ 1 in Tab.~\ref{TAB1}) already satisfies all indicators of successful intercalation mentioned above. Fig.~\ref{FIG2}(e) reveals the details of the central part of (c). New Pb superstructure spots appeared along the graphene direction. Panel (f) shows line profiles taken along the marked directions. The unusual broad shape, especially compared to the $(6\times6)_\text{SiC}$ spots, indicates the imperfect ordering of the superstructure, with an average periodicity close to the $(10\times10)_\text{Gr}$ supercell.
The next intercalation step included a repeated deposition of Pb followed by mild annealing for a long time (No.\ 2 in Tab.~\ref{TAB1}). The resulting diffraction pattern, shown in Fig.~\ref{FIG2}(g), revealed a subtle shift of superstructure spots towards $(00)$-spot. In direct comparison with the unchanged $(6\times6)_\text{SiC}$ spots, the line profiles in (h) now reveal a periodicity close to $(12\times12)_\text{Gr}$.
Fig.~\ref{FIG2}(d) shows a SEM image of the same area as in panel (a), but after various intercalation steps (No.\ 6 in Tab.~\ref{TAB1}). As obvious, the changes of the SEM contrast due to (further) intercalation are minor, but the appearance of additional contrast levels is related to successful intercalation.

In order to directly compare the resulting superstructure spot position after each preparation step, we extracted radial profiles from SPA-LEED measurements as shown in Fig.~\ref{FIG2}(e,g) by azimuthal averaging. This allows a precise measurement of the superstructure spot using the $(6\times6)_\text{SiC}$ spots as a reference as they are not expected to change their positions upon intercalation. We assume distortions of the inner section of the diffraction pattern to be minimal by alignment, so no additional corrections were applied. The results are summarized for the different intercalation protocols in Fig.~\ref{FIG3}. 
The positions of integer graphene supercells are marked for visual reference, the measured actual positions are listed in Tab.~\ref{TAB1}.  Concerning the Pb-induced spots, similar peak positions are found for similar preparation steps. The medium temperature annealing with previously deposited Pb (No.\ 1 and 3) align well with the $(10\times10)_\text{Gr}$ position, Pb deposition and mild annealing at 350~\degC\;(No.\ 2, 4 and 5) results in peak position ranging from $(11\times11)_\text{Gr}$ to $(13\times13)_\text{Gr}$. Surprisingly, the careful desorption of residual surface Pb clusters at 400 and 450~\degC\;(No. 6) yields the shortest periodicity, almost reaching $(9\times9)_\text{Gr}$.

\begin{table}[tb]
	\caption{\label{TAB1} Intercalation sequence containing several cycles of different deposition and annealing steps. The last column contains the more precise periodicity of the resulting intercalated phases related to the curve numbers in Fig.~\ref{FIG3}.}
	\centering
	\begin{tabular}{| c | l| l | c |}
		\hline
		No. in                & deposition:  & annealing:    & structure w.r.t. \\
		Fig.~\ref{FIG3} & $\theta_\text{Pb}$, \textit{T} & \textit{T}, time & graphene  \\ \hline \hline
		& 5~ML, 200~\degC & - & - \\	\hline
		1   & 20~ML, RT & 500~\degC, 5~min & $(10.2\times10.2)$ \\ \hline
		2   & 60~ML, RT & 350~\degC, 17~h & $(12.5\times12.5)$ \\ \hline
		& 20~ML, RT & 500~\degC, 5~min &   \\ \hline
		&           & 700~\degC, 5~min &   \\ \hline
		3   & \{20~ML, RT & 500~\degC, 5~min\}$\times4$  & $(10\times10)$ \\ \hline
		4   & 60~ML, RT & 350~\degC, 15~h & $(11.7\times11.7)$ \\ \hline
		5   & 60~ML, RT & 350~\degC, 65~h & $(11.7\times11.7)$ \\ \hline
		&           & 400~\degC, 10~min &   \\
		&           & 450~\degC, 10~min &   \\ \hline
		6   &           & 450~\degC, 10~min & $(9.2\times9.2)$ \\ \hline
	\end{tabular}
\end{table}

\begin{figure*}[tb]
	\begin{center}
		\centering
		\includegraphics[width=.9\linewidth]{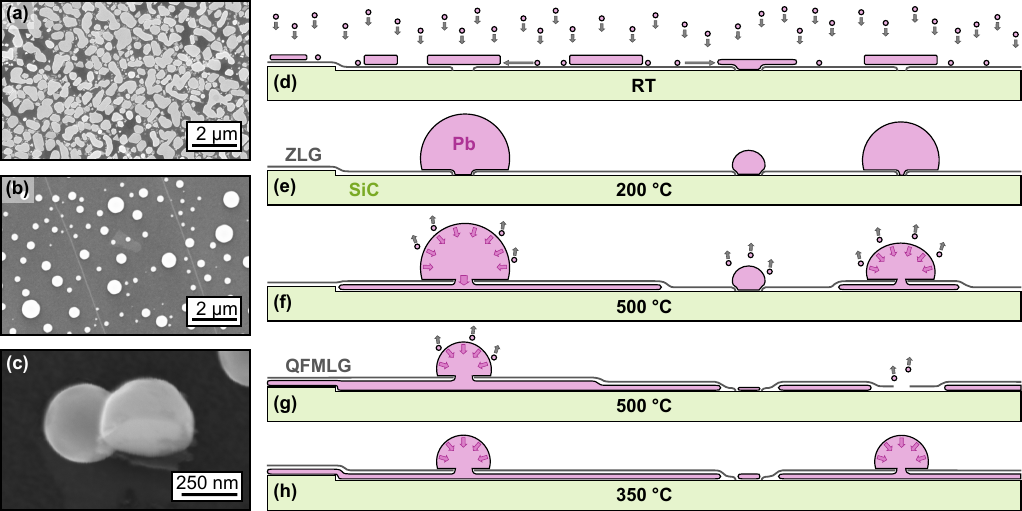}
        \caption{(a) SEM image of flat Pb(111) islands on MLG after deposition at RT. (b) SEM image of Wulff-shaped Pb clusters on an intercalated terrace observed after step 5 mentioned in Tab.~\ref{TAB1}. (c) Detail of Pb clusters. The right cluster was turned over by an STM tip, revealing its semi-spherical shape with a flat base. (d-h) The model of intercalation procedure based on observations at various stages: (d) Pb deposition, surface diffusion and nucleation at defects. (e) Formation of clusters at defect sites upon mild annealing. (f) Activation of Pb intercalation at elevated temperatures (e.g., 500~\degC) starting from the Pb clusters at defects. (g) Partial desorption from the clusters and deintercalation if the gradient of the chemical potential becomes too small. (h) Once the intercalation process is activated the intercalation yield can be increased if the temperature is slightly reduced. As deintercalation is prevented by non-desorbing Pb clusters, the interface relaxes at mild annealing temperatures. 
    \label{FIG4}}
	\end{center}
\end{figure*}

Following the different  intercalation protocols, the reversible transition between different periodicities becomes apparent (No.\ 1--4). The structure is barely affected by the repetition of same steps, i.e.,  mild long-time annealing (No.\ 4, 5),  but the intensities at medium temperature annealing without changing periodicity are increased (No.\ 1, 3). Intermediate steps at higher temperatures were found to boost the intercalation progress.

Thus, it can be concluded that a reversible transition occurs between a more densely packed phase, e.g,  $(10\times10)_\text{Gr}$, and a less dense structure, e.g., $(11\times11)_\text{Gr}$, as induced by intercalation cycles No.\ 2 and 1, respectively. The denser phase appears to be under considerable strain and may, therefore, be metastable. Its formation, triggered by brief heating to elevated temperatures, bears resemblance to a quenching process. In contrast, prolonged annealing at lower temperatures promotes the formation of more relaxed phases that resemble a disordered Pb layer, a structure that is thermodynamically favored due to increased entropy. In the following, in the context of STM measurements, we will present a model of the 2D interface structure that accounts for the observed variation of the reflex positions. 

\subsection{A model for the Pb intercalation}

 Compared to adsorption experiments, intercalation is more complex due to additional activation energies and diffusion constraints in reduced dimensions \cite{Han2023}.
Based on the systematic variation of the intercalation protocol, we propose the following model for the intercalation process. So far, direct intercalation, especially of heavier elements, from the gas phase has not been successful. Therefore, the material was deposited on ZLG at temperatures below the actual intercalation threshold. This already indicates that the energy barriers for intercalation are significantly higher than those for desorption. 

As shown in Fig.~\ref{FIG4}(a) and (d), the Pb atoms deposited at room temperature are highly diffusive and form epitaxial Pb(111) island on ZLG as apparent from characteristic diffraction spots seen in LEED  \cite{Chen2020, Gruschwitz2021, Schoelzel2024}. In case that the initial surfaces contain fractions of MLG, the Pb islands form preferentially on ZLG, showing the higher surface energy \cite{Liu2015, Hupalo2011}. 

At slightly elevated temperatures, the Pb(111) islands melt, coalesce and finally form semi-spherical droplets as observed by SEM after intercalation without complete desorption of residual Pb. After repeated desorption steps shrinking clusters are found at unchanged positions suggesting the pinning to defect sides of the ZLG from where intercalation sets in (Fig.~\ref{FIG4}b and e). The formation of molten Pb spheres is consistent with surface energy minimization and weak substrate interaction \cite{Metois1989, Zhang1997}. This observation contrasts with a recent LEEM study, where the disappearance of Pb(111) spots was attributed to melting and the formation of a uniform liquid overlayer \cite{Schoelzel2024}. 
 
Intercalation sets in at higher temperatures. In our study, annealing to  500~\degC\; activates/opens the defects for intercalation, e.g., by weakening bonds to the substrate. Thereby, the surface tension of the spherical Pb clusters facilitates the intercalation by applying pressure on the intercalation front (Fig.~\ref{FIG4}). Such a  mechanism may also occur in the self-propelled intercalation observed for Ga even at room temperature at RT \cite{Wundrack2021}.

During the intercalation process, the gradual increase in the cluster curvature facilitates continued intercalation. The resulting chemical potential gradient across the interface toward the diffusion front helps to compensate for the growing kinetic limitations of Pb atom diffusion. However, due to the gradually increasing curvature, the desorption rate of Pb atoms from the clusters should also increase, causing a disproportionate loss of the available Pb reservoir.  Therefore, after the initiation of the intercalation process, a reduction of the intercalation temperature, e.g., as sketched in Fig.~\ref{FIG4}(h), may be helpful in order to have largely intercalated areas. The balance between initial cluster size and intercalation temperature also determines the amount of intercalated Pb atoms. As discussed in the context of Fig.~\ref{FIG3}, the different intercalation protocols result in slightly varying periodicities seen in electron diffraction. As we will show and discuss in more detail in the following section using STM, these periodicities, as well as different structures, are indeed associated with different Pb interface densities. 

Once the Pb reservoir is exhausted during annealing at 500~\degC, deintercalation sets in. The "gates" that previously facilitated access to the interface now act as pathways for Pb atoms to escape. Depending on the degree of deintercalation, subsequent deposition and annealing steps may still promote further intercalation. However, if deintercalation is too extensive, resulting in the formation of a stable (or even more stable) ZLG configuration, intercalation may no longer proceed at the original defect sites. Some of our samples showed no improvement upon further deposition and annealing, particularly after high-temperature treatments (e.g., 700~\degC\; for 5~min or 1000~\degC\; for 15~s). Nevertheless, samples subjected to such high-temperature steps exhibited well-ordered Pb superstructure spots and demonstrated strong stability against additional annealing.

In the case of Pb, the intercalation yield is critically dependent on the initial defect density of the ZLG.
The intercalation of high-quality samples produced by polymer-assisted sublimation growth (PASG) is often more difficult and shows a rapid intercalation along terraces and a delayed intercalation across substrate steps \cite{Schoelzel2024}. Our Ar-atmosphere grown sample hosts much taller, bunched substrate steps. Intercalation dominantly proceeds along these and is negligible across terrace steps.

\section{B. Results and Discussion: the structure of the intercalated phase}

\subsection{Different 2D Pb phases studied by low energy electron and scanning tunneling microscopy (LEEM, STM)}

\begin{figure}[tb]
\begin{center}
	\includegraphics[width=.6\columnwidth]{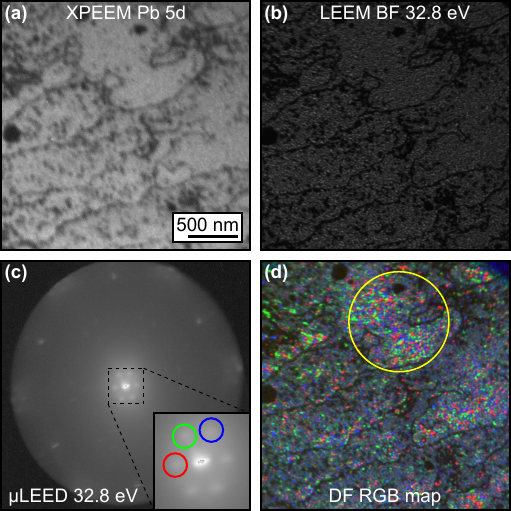}
	\caption{(a) XPEEM image based on the Pb 5d level reveals that the majority of the surface was successfully intercalated. Photon energy $h\nu=150$~eV and electron kinetic energy 47~eV. (b) LEEM bright field image of the same area, revealing local inhomogeneities within the intercalated phase.  (c) µLEED  pattern  (field of view 100~µm). The marked spots resemble the quasi-$(10\times10)_\text{Gr}$ reconstruction spots as shown in context of Figs.~\ref{FIG1} and \ref{FIG2}. (d) Superimposed dark field maps of the three spots marked in (c). Grey and colored areas denote the formation of Pb interface structure with and without a six-fold symmetry, respectively. Black regions match to the non-intercalated areas seen in (a).  All LEEM and LEED measurements were carried out at $E=32.8$~eV. \label{FIG5}
    }
\end{center}
\end{figure}

In connection with the Pb intercalation, different  phases were reported in the literature. In addition to the hexagonal phase, so-called stripe phases were also found \cite{ Yurtsever2016, Gruschwitz2021, Hu2021, Schaedlich2023, Vera2024, Wang2025}.
To gain deeper insight into the structural origin of the previously mentioned reconstruction spots, microscopy experiments were conducted. We present here dark-field (DF) mappings using LEEM to visualize the spatial distribution of the reconstructed species.

The intensity mapping of Pb 5d photoelectrons at $E_\text{kin} = 47$~eV, shown in Fig.~\ref{FIG5}(a), again indicates a continuous Pb layer across the terraces, interrupted only by minor defects and step edges. These interruptions largely coincide with the inhomogeneities observed in the LEEM bright-field image at 32.8~eV shown in panel (b).
Local LEED measurements performed with a 100~µm field of view clearly reveal Pb-induced superstructure spots (see Fig.~\ref{FIG5}c), similar to those observed in the SPA-LEED images in Fig.~\ref{FIG1}. The inset highlights the positions of the DF apertures, aligned with three adjacent superstructure spots. The corresponding intensity maps, color-coded in red, green, and blue, are overlaid in Fig.~\ref{FIG5}(d). While many terraces show only faint colored speckles on a gray background, others, such as the yellow-circled region, appear vividly colored, indicating the presence of well-ordered stripe domains extending over micrometer-wide areas. Individual domains can reach sizes of up to 100~nm. Consequently, the gray regions correspond to intercalated phases exhibiting six-fold symmetry.
Previously, stripe phases-often coexistent with a hexagonal phase-where only observed on small scales by STM \cite{Yurtsever2016, Hu2021, Gruschwitz2021, Ghosal2022}. An atomistic model explaining both types of superstructures and their relation has not yet been proposed. Recently, strained Pb structures were considered as the origin of the stripe phase \cite{Vera2024, Wang2025}. In the following, we will show that these two structures can indeed coexist and are closely related.

\begin{figure}[tb]
\begin{center}
	\includegraphics[width=.6\linewidth]{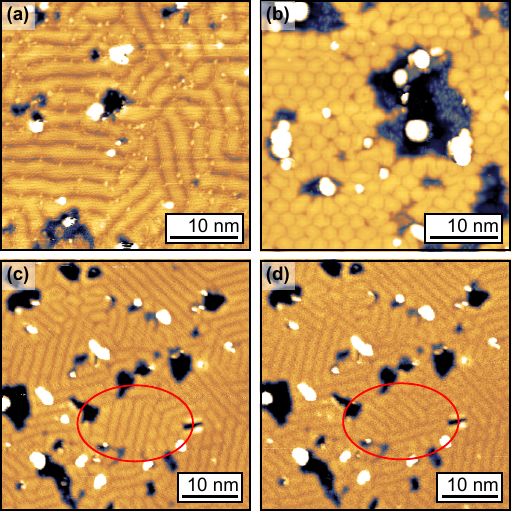}
	\caption{STM images showing various 2D Pb interface phases  at the SiC graphene interface. (a)  2-fold symmetric, three domain stripe phase. (b) hexagonal phase exhibits deformations in close vicinity to defects and non-intercalated areas. (c) Often mixed phases with one dominant phase are observed. (d) During continuous scanning local rearrangements of grain boundaries, highlighted by the red circle, can are observed. Tunneling parameters: (a) LHe, 1~V, 50~pA; (b) LN$_2$, 2~V, 200~pA; (c) RT, 1~V, 150~pA; (d) RT, 1.2~V, 150~pA.  \label{FIG6}
    }
\end{center}
\end{figure}

In addition, we analyzed the 2D Pb phases with STM.
Fig.~\ref{FIG6}(a) and (b) display semi-ordered stripe and hexagonal phases, respectively. The local structure appears closely linked to the presence of defects. Pb superstructures surrounding defects and the edges of non-intercalated (dark) regions align these irregularities with adjacent supercells. Stripes frequently attach perpendicular to larger defects, suggesting a preferred direction for strain relief. Strain along the stripe direction can also be mitigated by introducing additional kinks in the domain walls. As more line defects are added, the stripe pattern gradually evolves into a hexagonal network of domain walls.

Fig.~\ref{FIG6}(c) and (d) demonstrate a spontaneous phase transition observed during continuous STM scanning of the same area. These measurements not only support the LEEM results discussed above and confirm the coexistence of two distinct Pb interface structures, but also highlight the low activation energy required for structural transformations, i.e. reorientation of stripes and formation of stripes from hexagonal phases. 
Comparable transformations have been reported in Pb-intercalated monolayer graphene (MLG) \cite{Hu2021}. Such structural changes imply a degree of flexibility in the interfacial Pb layer, particularly for Pb atoms that are weakly bonded to the substrate. Given the rapid intercalation at 350~\degC \cite{Schoelzel2024}, local relaxation events cannot be ruled out. However, the phase transition is more likely triggered by a high electric field induced by the STM tip during scan instabilities.
The abrupt reorientation of stripe domains observed here suggests a collective phase transition involving interconnected Pb regions. Similar rapid, collective self-organisation and superdiffusion have been reported in incommensurate systems like Pb on Si(111) \cite{Yakes2004, Man2013} and are often understood within a generalized Frenkel-Kontorova model \cite{Huang2012}. Epitaxial layers close to the commensurate-incommensurate phase transition are known to exhibit a wide range of superstructures with low barriers for transitions between them \cite{Bak1982}. Although incommensurate layers on surfaces were investigated intensively, such lattices at interfaces are mostly unexplored.
In the following, we present atomistic models demonstrating that both structural motifs arise from strain effects accompanied by characteristic grain boundary formation.

\subsection{Atomistic models for striped and hexagonal phases}

\subsubsection{Striped phase structure}

\begin{figure}[htb]
\begin{center}
	\includegraphics[width=.5\linewidth]{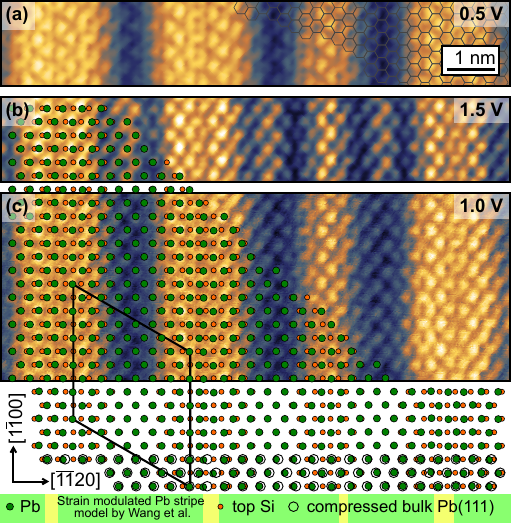}
	\caption{High resolution STM scans of the stripe phase. (a) At low bias voltages (0.5~V) the honeycomb symmetry of the graphene on top is most apparent. The overlaid graphene structure emphasizes the alignment of the edges of the stripes with the graphene zigzag direction. (b) At high bias voltages (1.5~V) the SiC substrate structure is imaged.  However, the appearance is modulated by the adjacent layers. (c) Medium bias voltages (1.0~V) are most sensitive to the Pb-induced distortion of the substrate appearance. In the lower half we superimposed the strain modulated stripe phase model (green circles) proposed by Wang et al. \cite{Wang2025}. The colorbar below the model indicates regions of the non-altered sequence of Pb positions according to the model (green) and inserted columns for continued agreement with the measurement (yellow). The open circles show the similarity of the periodically strained model to an uniaxially compressed bulk-Pb(111) lattice. All measurements were performed with 25~pA tunneling current and at $T=4$~K.
    \label{FIG7}
    }
\end{center}
\end{figure}

In order to investigate the stripes seen in  Fig.~\ref{FIG6} in more detail, we have performed low-temperature STM experiments at various bias voltages. Fig.~\ref{FIG7} presents the detailed structure of such structures at different bias voltages. While measurements at low bias (Fig.~\ref{FIG7}a, 0.5~V) are more sensitive to the graphene layer on top, high bias voltages (b, 1.5~V) are able to image mainly the  SiC surface. Especially at intermediate voltages (panel c, 1.0~V) the presence of a Pb layer at the interface becomes apparent as a distortion of those lattices. The bias-dependent STM images are one qualitatively similar to previous measurements \cite{Ghosal2022}. However, for a quantitative analysis, we had to assume that at least two intercalated Pb layers are necessary to account for the observed structures and moiré features. The successful intercalation of multilayers was indeed confirmed by TEM (cf. SI in Ref. \cite{Vera2024}).

Recently, the Pb interface structure was discussed in the context of strain relaxation, allowing for the stabilization of Pb grains rather than extended highly compressed Pb$(1\times1)_\text{SiC}$ structures \cite{Vera2024}. The extension to larger unit cells, e.g.,$(10\times 10)_\text{Gr}$,  yields a strain-based model of the stripe phase comparable to experimentally observed periodicities \cite{Wang2025}. Annealing an incomplete Pb$(1\times1)_\text{SiC}$ layer at the interface results in a striped structure aligned along the zigzag direction of graphene. These stripes arise from the precise alignment of Pb rows with the top-layer Si atoms along the $(1\bar{1}00)$ planes and a periodic modulation of Pb–Pb spacing along $[\bar{1}\bar{1}20]_\text{SiC}$. This modulation reflects a balance between in-plane Pb interactions and substrate pinning via unsaturated Si dangling bonds, leading to a strain variation relative to a relaxed Pb(111)-like layer. Notably, the stripe periodicity is a low-order commensurate match to a uniaxially compressed Pb(111)-like structure along the $[1\bar{1}00]$ direction, as illustrated by the lattice of open circles in Fig.~\ref{FIG7}(c), which exhibits a horizontal periodicity of 3.51~\AA.
Assuming an isochoric transformation, the corresponding unstrained Pb(111)-like layer would have a lattice constant of 3.29~\AA, close to the values calculated for freestanding Pb(111)-like monolayer by Han et al. \cite{Han2024} and the lattice constant of $\alpha$-Pb on Si(111) \cite{Chan2003}. The most compressed regions align nearly with the $(1\times1)_\text{SiC}$ of the substrate, while a gradual transition from compressive to tensile strain positions Pb atoms at Si bridge sites. The mismatch of the strained Pb rows compared to the equally spaced circles nicely resembles the calculated non-linear offset of Pb atoms w.r.t. Si top positions calculated by Vera et al.\ \cite{Vera2024}. The stripe model unit cell SiC$(5\sqrt{3}\times5\sqrt{3})R30^\circ$ by Wang et al.\ \cite{Wang2025} outlined by the black rhombus in Fig.~\ref{FIG7}(c), is overlaid on the STM measurements. We find good agreement between the model and the measurement over one stripe period. 
However, to reproduce the full extent of the experimentally observed stripes, it was necessary to locally modify the sequence by adding or removing strained Pb columns. The color bar below the model indicates unaltered regions (green) consistent with the original model and manually adjusted regions (yellow).

Although for the striped phase hints towards  grain boundaries were observed \cite{Yurtsever2016, Ghosal2022}, the strain model discussed here rather suggests a gradual transition. In the following, we will show that the striped phases can be considered as building blocks for the hexagonal phase. However, in this case, the compressive/tensile strain profile can no longer be gradually relaxed, leading to the formation of grain boundaries (GB) at least at low temperature. It turns out that at elevated temperatures the hexagonal phase form extended $\alpha$-Pb structures.  

\subsubsection{Structure of the hexagonal (bubble) phase}
\begin{figure}[htb]
	\begin{center}
		\includegraphics[width=.45\columnwidth]{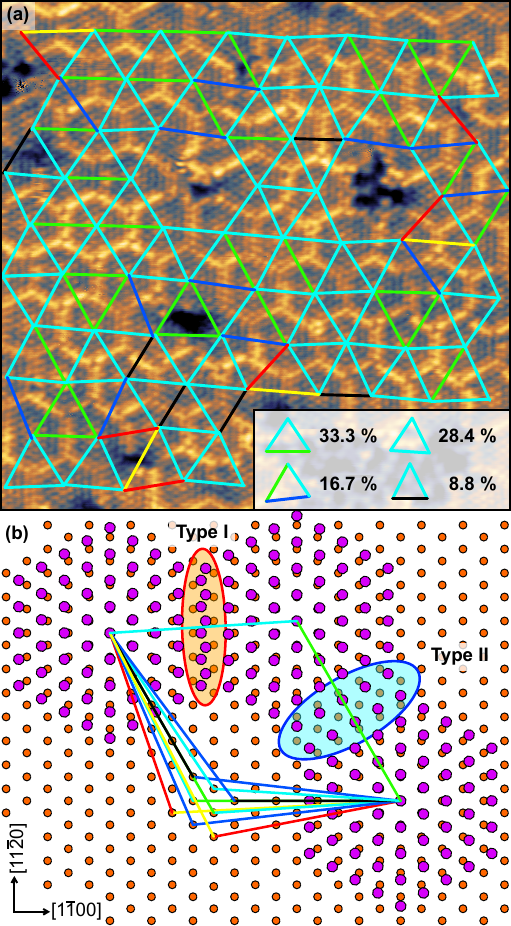}
		\caption{Analysis of grain boundary structures formed within the hexagonal phase from a high resolution STM image (1~V, 0.1~nA, 4~K). The graphene lattice is visible at these tunneling conditions. (a) Colors represent different distances and angular orientations between neighboring grain centers. The cyan and green bars correspond to the most frequently observed configurations (see inset), while the others are typically found near defects. (b) Each colored bar corresponds to a specific GB type, defined by the shift direction relative to the underlying symmetry axes. The model structure for the most common arrangement is shown, along with possible reconstructions corresponding to the other colored bars. Depending on the relative shift of the centers, either grain boundaries  with mirror (Type I) or  glide mirror symmetry (Type II) are formed. 
			\label{FIG8}
		}
	\end{center}
\end{figure}

\begin{table}
	\caption{\label{TAB2} Characteristic Pb grain distances $d_\text{SiC}$ and $d_\text{Gr}$ w.r.t the SiC and graphene (Gr) lattices, respectively.  The results are obtained from the analysis of the STM image in Fig.~\ref{FIG8}(a).  The colors refer to those used in the STM image. The occurrence of $d_\text{SiC}$ and the  type of GB shown in Fig.~\ref{FIG8}(b) are mentioned as well. }
	\centering
	\begin{tabular}{| c | c | c | c | c |}
		\hline
		Color  & $d_\text{SiC}$                 & $d_\text{Gr}$        & Occur.  & GB \\ \hline \hline
		cyan   & $\sqrt{61}R(30\pm3.7)^\circ$  & $9.77R\pm3.7^\circ$  & 64.5~\% & I \\ \hline
		green  & $5\sqrt{3}R30^\circ$          & $10.84R0^\circ$      & 20.0~\% & II \\ \hline
		blue   & $2\sqrt{17}R(30\pm6.6)^\circ$ & $10.32R\pm6.6^\circ$ & 6.7~\%  & II \\ \hline
		black  & $4\sqrt{3}R30^\circ$          & $8.67R0^\circ$       & 3.3~\%  & II \\ \hline
		red    & $3\sqrt{7}R(30\pm10.9)^\circ$ & $9.94R\pm10.9^\circ$ & 3.3~\%  & I \\ \hline
		yellow & $\sqrt{91}R(30\pm3)^\circ$    & $11.94R\pm3^\circ$   & 2.2~\%  & I \\ \hline
	\end{tabular}
\end{table}

Previously, a model for the hexagonal phase (also called bubble or mottled phase) based on a symmetrized $\alpha$-Pb phase on Si(111) was  established \cite{Schaedlich2023}. This model assumes a full coverage of Pb with grains hosting Pb structures exceeding the SiC lattice constant, necessarily leading to heavy grain boundaries. Although such dense structures are expected to buckle, no clear indications have yet been observed \cite{Schaedlich2023}.  
We present in the following a refined model for the formation of the hexagonal phase by considering also the intercalation process itself.

To gain deeper insight into the hexagonal phase formation, we performed a detailed analysis of high-resolution STM measurements, focusing on the relative grain arrangement and resulting grain boundaries. In Fig.~\ref{FIG8}(a), we superimpose vectors connecting the central pinning sites of neighboring grains. The pinning sites were chosen considering the general grain symmetry and slightly increased intensity of the inner seven atoms. As suggested by recent calculations, these central Pb atoms are expected to align with topmost Si atoms of the substrate (T1-sites) \cite{Unigarro2025}. Vectors of the same color indicate equivalent arrangements in terms of distance and relative orientation. Their different lengths and orientation w.r.t.\ the SiC and graphene lattice are summarized in Tab.~\ref{TAB2}. 
Cyan and green arrangements appear most often and reveal roughly a 3:1 ratio. The inclusion of the green vector, corresponding to the 5$\sqrt{3}$ vector shown in Fig.~\ref{FIG7} also explains, compared to the idealized scenario discussed by Sch\"adlich et al.\ \cite{Schaedlich2023}, the symmetric appearance of superstructure spots rather than a angular splitting of a single periodicity.  Close to defects,  also varying appearances were found for the same neighbor vectors. As it turns out, the variability of the grain boundary structures is increased in the vicinity of defect sites. 
In Fig.~\ref{FIG8}(b), we schematically show  possible hexagonal phase arrangements according to those vectors. This results in two distinct types of grain boundaries. In case the  hexagons align along a single SiC direction, as indicated by the green or black vectors, mirror-symmetric boundaries (Type II) emerge. In all other cases, the boundaries exhibit glide-plane symmetry (Type I).
Most importantly, the symmetries derived from these fundamental vectors closely match the various Pb-induced reconstructions discussed in Tab.~\ref{TAB1} and Fig.~\ref{FIG3}. Minor variations in the intercalation protocol and subtle differences in defect types within the ZLG appear to enable the formation of slightly different grain boundary structures.

\begin{table}
	\caption{\label{TAB3} Moiré structures of hexagonal Pb lattices (symmetrized $\alpha$-Pb and Pb(111)) on SiC(0001) and their coverages. The strain applied to reach commensurability can be  tensile (+) or  compressive (-).}
	\centering
	\begin{tabular}{| l | c | c | c | c |}
		\hline
		$a_\text{Pb} $  & Pb $(m\times m)$   & SiC $(n\times n)$   & Strain  & Coverage \\
		\hline \hline
		\multirow{3}{4em}{3.33~\AA\, $\alpha$-Pb on Si(111)}  & 12 & 13 & $+0.20$~\% & 0.85 \\
		& 13 & 14 & $-0.39$~\% & 0.86 \\
		& 14 & 15 & $-0.91$~\% & 0.87 \\ \hline
		\multirow{4}{4em}{3.51~\AA\, bulk Pb(111)}          & 12 & 14 & $+2.31$~\% & 0.73 \\
		& 13 & 15 & $+1.23$~\% & 0.75 \\
		& 14 & 16 & $+0.28$~\% & 0.77 \\
		& 15 & 17 & $-0.55$~\% & 0.78 \\ \hline
	\end{tabular}
\end{table}

Considering the 'tiles' formed by the vectors, most commonly two cyan vectors are included and the third one is either a green or cyan vector. Both can realize a full tiling themselves, with the equilateral tiles forming the previously suggested model for the hexagonal phase \cite{Schaedlich2023, Schoelzel2024}. The isosceles tiles can only arrange in a stripe-like structure along the green vector, close to the ones shown in Fig.~\ref{FIG7}. The unit cells form equilateral (purely cyan) or isosceles triangles (cyan and green) which are slightly rotated with respect to SiC and would follow the  $\big(\begin{smallmatrix} 9 & 4 \\ 5 & 9 \end{smallmatrix}\big)$ and $\big(\begin{smallmatrix} 10 & 5 \\ 5 & 9 \end{smallmatrix}\big)$ Woods notation, respectively. 
Considering the observed vectors and their linear combinations extending to second-nearest neighbors (Fig.~\ref{FIG8}b), a notable number align with commensurate positions. This higher-order symmetry corresponds to a $(14\times14)_\text{SiC}$ supercell. Tab.~\ref{TAB3} lists various moiré scenarios involving hexagonal Pb MLs with the two most likely lattice constants. Each scenario yields slightly different Pb coverages defined with respect to the top Si atom density. Nevertheless, highest coverages with lowest strain are obtained with $\alpha$-Pb phases, which are the building blocks for the GB model \cite{Schaedlich2023}.

\begin{figure}[tb]
\begin{center}
	\includegraphics[width=.9\columnwidth]{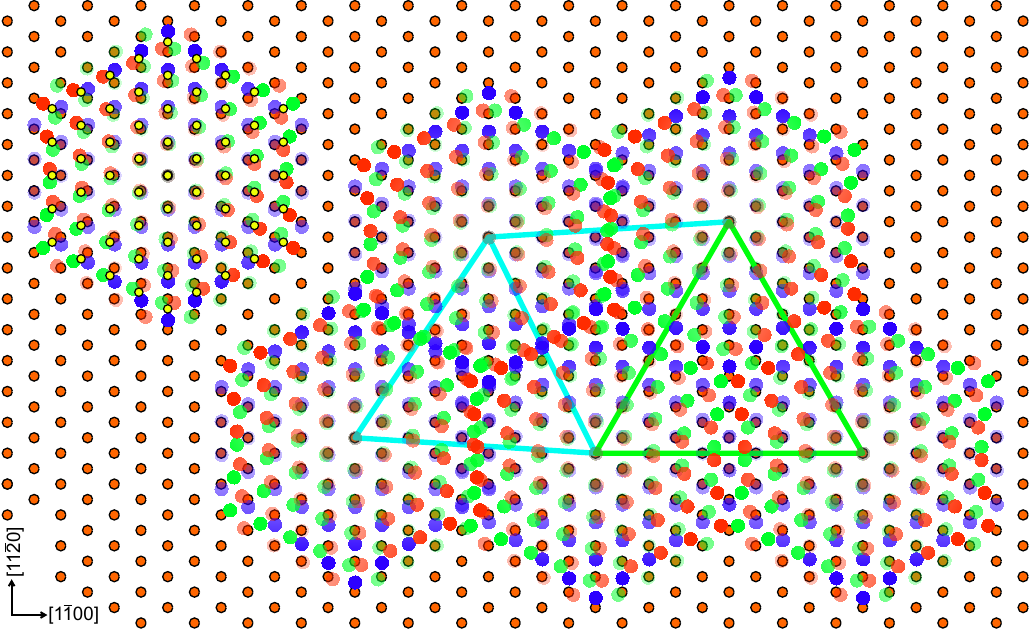}
	\caption{Hexagonal phase constructed from three $120^\circ$ rotated and overlaid stripe models according to Wang et al. \cite{Wang2025}. Each color represents one stripe orientation, the colors fade towards the compressive strain center of each stripe (cf. Fig.~\ref{FIG7}). The directions of the maximum compression planes for the red, blue  and green  stripes are $[01\bar{1}0]$, $[10\bar{1}0]$ and $[\bar{1}100]$, respectively. The single hexagon in the top left is superimposed by a $\alpha$-Pb-like lattice (yellow circles). It coincides well with the center of the three possible adsorption sites according to the uniaxially compressed stripe model, supporting the isochoric transformation due to strain. The five hexagons are aligned according to the most commonly observed neighboring vectors (cyan, green).
	\label{FIG9}
	}
\end{center}
\end{figure}

The hexagonal phase can also be understood and constructed via the strain model, discussed in Fig.~\ref{FIG7} and based on recent DFT calculations \cite{Wang2025}. The transition to a radial strain distribution, characteristic of the hexagonal phase, is obtained by superimposing three stripe phases, each rotated by $120^\circ$. The resulting structure, shown in the top left of Fig.~\ref{FIG9}, features three color-coded sublattices, with transparency indicating the strain gradient, increasing toward the highly compressed rows. Each SiC lattice site corresponds to three potential Pb adsorption sites. To achieve radial symmetry for a single hexagon, these adsorption sites must be symmetrized. When all three sites associated with a single top-Si atom are considered, their centroid aligns closely with the $\alpha$-Pb lattice positions, marked by small yellow dots. 
Moreover, the model proposed in Fig.~\ref{FIG9} also may help to better understand the domain wall structure. Similar to the model proposed in the context of Fig.~\ref{FIG8}(b), the hexagonal phase, formed from the striped phase, reveals GB structures with similar symmetries, i.e., type~I and type~II for the cyan and green unit cell vectors, respectively. The model can also plausibly explain possible detailed structures of the dislocation. By occupying the quasi-overlapping positions with a Pb atom, an almost continuous transition between the grains can be achieved. 
As we will show below, mild thermal annealing indeed appears to allow for the formation of this structure, i.e., the quasi-$(10 \times 10)_{\text{Gr}}$ reconstruction vanishes, and only the lattice spots of the $\alpha$-Pb phase remain visible.

\subsubsection{Formation of the hexagonal phase}

\begin{figure*}[tb]
\begin{center}
	\includegraphics[width=.9\linewidth]{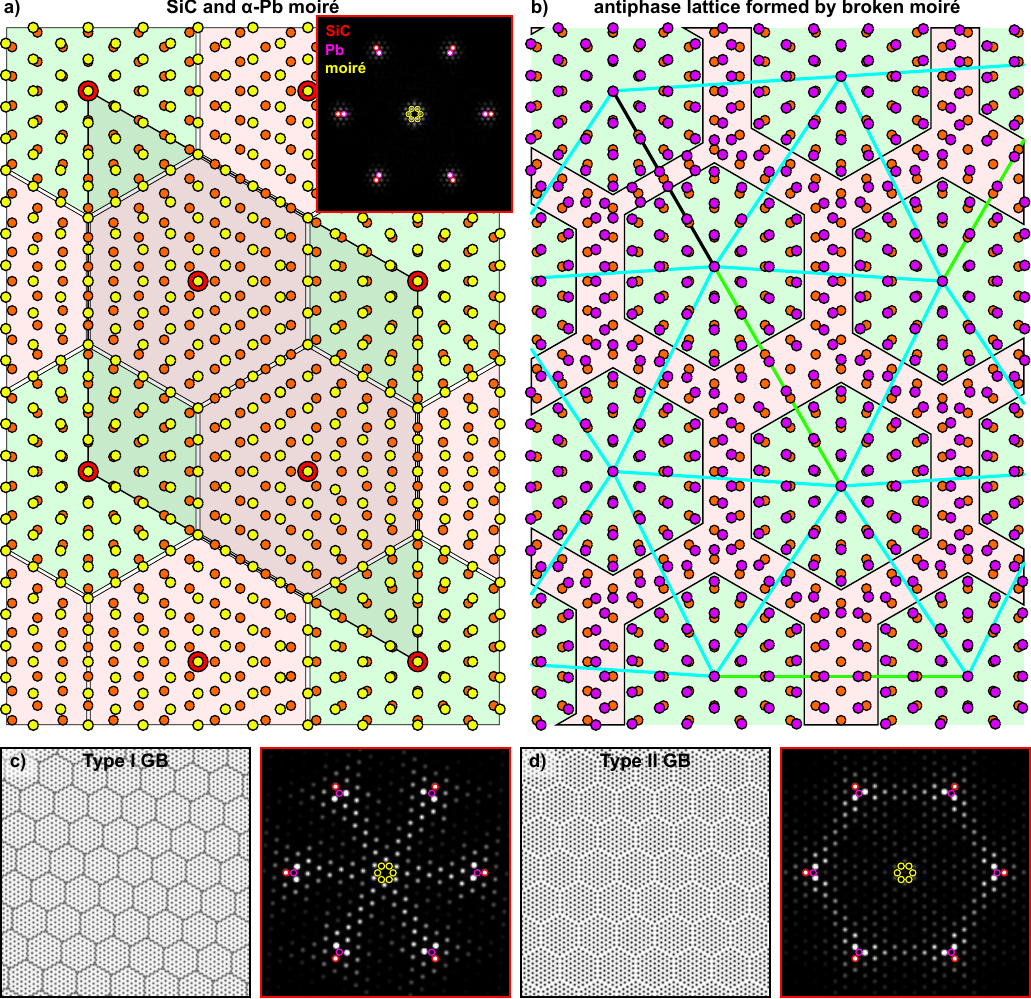}
	\caption{Model of hexagonal phase formation due to interface reconstruction upon sample cooling after annealing. a) hexagonal Pb layer (yellow, $\alpha$-Pb) of 3.33~\AA\, lattice constant superimposed to the Si atoms of SiC(0001). The displayed FFT reveals typical moiré spots (yellow circles) as well as first-order spots of the Pb lattice (magenta circles). In real space the moiré pattern can be separated into areas of small lattice sites mismatch (green) and regions of large accumulated lattice mismatch, surrounding the symmetric out-of-phase positions. b) Including substrate interaction results in local pinning of Pb atoms to high symmetry sites to T1 positions. The red out-of-phase hexagons experience a rigid shift to one of the neighboring T1 sites, resulting  in a periodic arrangement close to the observed $(10\times10)_\text{Gr}$. For the sake of simplicity, we used the GBs classified in Fig.~\ref{FIG8}b). (c,d) Hexagonal phases with type I and type II GBs, respectively, and their corresponding FFTs showing a complete suppression of the Pb and moiré first order spots but  appearance the quasi-$(10\times10)_\text{Gr}$ spots.\label{FIG10}}
\end{center}
\end{figure*}

Fig.~\ref{FIG10} illustrates the proposed relaxation mechanism of a hexagonal Pb layer, from an unpinned configuration without substrate interaction (panel a) to a hexagonal phase model stabilized by pinning at T1 sites (panel b). In the moiré model, red circles mark Pb atoms at high-symmetry points (T1 and H3). Green hexagons indicate in-phase regions where Pb atoms are only slightly displaced from underlying Si top sites, while red hexagons mark atoms significantly offset, with the central Pb atom at a perfect out-of-phase position (H3). The inset in Fig.~\ref{FIG10}(a) show the Fourier transform of this moiré lattice revealing  clearly first-order Si, Pb, and moiré spots.
When coupling to the substrate is considered, these central Pb atoms shift toward nearby Si top sites, forming  hexagons equivalent to the green hexagons in Fig.~\ref{FIG10}(a). Depending on the direction of this shift, either cyan or green neighbor vectors emerge.
This spontaneous symmetry breaking leads to an extended anti-phase lattice, with grain boundaries formed by Pb atoms outside the  centers of the hexagons once accumulated strain overcomes the pinning force (b, red regions). The resulting quasi-periodicity of in-phase (green) hexagons produces broad superstructure diffraction features. However, due to the anti-phase arrangement, first-order Pb and moiré spots from the commensurate lattice in (a) are completely suppressed.
Fig.~\ref{FIG10}(c,d) shows model lattices formed with type I or type II GBs, respectively. The hexagonal phase consists of anti-phase $\alpha$-Pb grains so that the first order Pb spots are vanishing. Also, instead of the moiré spots (yellow circles), the quasi-$(10\times10)_\text{Gr}$ spots show up.  The calculated Fourier transform reveals different intensity distributions depending on the GB type. Surprisingly, the light type II GBs result in a significant reduction of the intensity of the first-order superstructure spots. A more detailed analysis of the superstructure spots intensity distribution can provide more details about the domain wall type and distribution \cite{Zeppenfeld1988, Timmer2017}.

\begin{figure}[th]
\begin{center}
	\includegraphics[width=.8\linewidth]{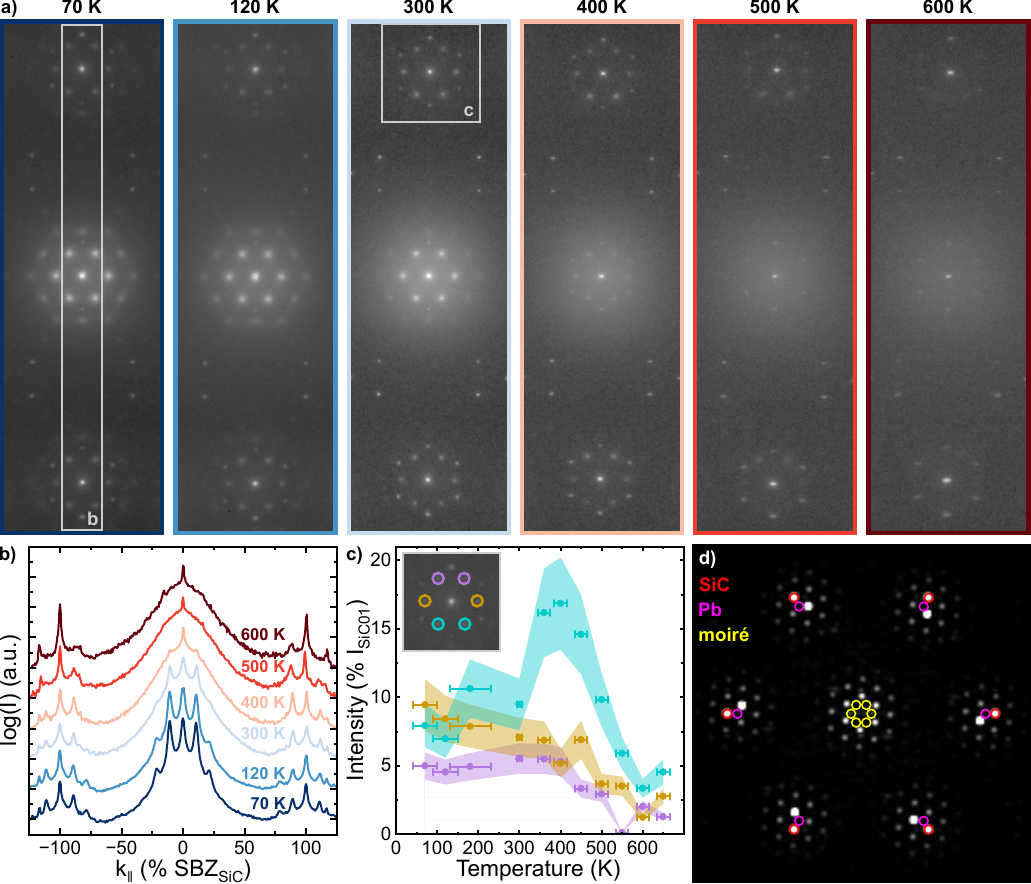}
	\caption{a) Section of normalized SPA-LEED images along $k_{||\text{, SiC}}$ taken at $E=168$~eV at various temperatures. The box in the 70~K measurement marks the direction and averaging width of the extracted line profiles plotted in b). A constant offset is added for better visibility. While the quasi-$(10\times10)_\text{Gr}$ reflexes completely vanish at 500~K, two Pb diffraction spots reach maximum intensity relative to the SiC(10) reflex at 400~K. The intensity plot in c) reveals a significant difference in the temperature dependence of the Pb spots around SiC(10). The intensity maxima can be associated with first-order diffraction spots of one $\alpha$-Pb domain rotated by $3.7^{\circ}$ w.r.t. $k_{||\text{, SiC}}$. The Fourier transform of a corresponding artificial lattice is shown in d. \label{FIG11}
    }
\end{center}
\end{figure}

Finally, we present in Fig.~\ref{FIG11}(a,b) a sequence of SPA-LEED images and profiles, respectively, recorded at different substrate temperatures. The analysis of the integral intensities of Pb-induced reconstruction spots above the background reveal a different temperature dependent behavior of the two Pb spots marked by cyan circles. At an elevated temperature of around 450~K, the quasi-$(10\times10)_\text{Gr}$ spots, except for these two, vanish. Moreover, these remaining spots in the vicinity of the first-order SiC spots are closest to the first-order spots of the $\alpha$-Pb lattice (indicated in the Fourier transforms by purple circles). In the context of the hexagonal phase model shown in Fig.~\ref{FIG9}, it appears that at these elevated temperatures the hexagons "merge" to form extended $\alpha$-Pb domains. However, these domains remain misaligned with respect to the SiC substrate. From the experiments, we deduce a misalignment of approximately $3.7^\circ$. Indeed, the Fourier transform of such a lattice, shown in panel (d), nicely reproduces our experimental finding for one domain. The annealing may soften the bonds to the Si atoms. Nevertheless, the thermal expansion of the Pb lattice should also be considered. From the model discussed in Fig.~\ref{FIG9} the mismatch of the two hexagons for type I GB is around 4\% of the $\alpha$-Pb lattice constant. This would demand an thermal expansion coefficient of around 4-6$\times10^{-5}$ \cite{Zhang2007}. A comparable scenario was observed in case of Cu deposited on Si(111) forming a quasi-$(5\times5)_\text{Si}$ reconstruction \cite{Zegenhagen1992}. The mild annealing can be understood as a different balance between in-plane Pb interaction and Pb-substrate interaction, therefore leading to different reconstructions.

\section{Conclusions}

In summary, we conducted a detailed intercalation study of Pb beneath the ZLG on SiC(0001) and investigated the intercalation process and resulting interface structure using a multiscale microscopy approach (SEM, LEEM, SPA-LEED, STM). Our results show that the interface reconstruction of Pb can be tuned, within limits, by varying the details of the intercalation protocol. Based on our experiments, we propose an intercalation model that accounts for the different periodicities observed in the interface layer. The interplay between the size-dependent chemical potential of the Pb clusters and the activated yet spatially constrained diffusion process appears to be delicate and may partially explain the variations reported in the literature. Overall, both the thermodynamics and kinetics of the process are strongly influenced by the defect landscape of the ZLG.

In the second part, we investigated the atomic structure of the intercalated Pb interface layer. Both stripe and hexagonal-like patterns were observed. Based on recent calculations, we propose a uniaxially strained Pb(111) model, which aligns well with the striped structures observed in STM. Furthermore, the superposition of this model helps rationalize the formation of the hexagonal phase. In this case, the compressed Pb(111) layer becomes effectively isotropically strained, and the centers of the hexagons exhibit the crystal structure of the $\alpha$-Pb phase known from Si(111), consistent with recent measurements \cite{Schaedlich2023}.

Since intercalation occurs at elevated temperatures, we speculate that the hexagonal phase originate from a relaxed $\alpha$-Pb structure beneath the graphene. A moiré scenario between the two 2D layers comes along with too many energetically unfavorable adsorption sites. To realize favorable T1 adsorption sites, the Pb monolayer forms hexagonal phases separated by grain boundaries. Depending on the cooling procedure and the effective Pb coverage, hexagons of slightly varying sizes and separations are formed-consistent with our observations.

\section{Materials and Methods}

Zero layer graphene (ZLG) samples were grown on on-axis, n-doped 6H- or 4H-SiC(0001) substrates (6H: SiCrystal GmbH, 4H: Cree) by thermal decomposition in Ar atmosphere. Etching the substrates in a hydrogen atmosphere (6H: 800~mbar, 1550~\degC; 4H: 1000~mbar, 1425~\degC) yields clean, atomically flat SiC terraces. Subsequent annealing in Ar atmosphere (6H: 800~mbar, 1460~\degC; 4H: 1000~mbar, 1475~\degC) graphitized the surface, accompanied by the stabilization of 3-4~µm wide terraces due to step bunching. The coverage of the ZLG is controlled by the annealing time. Further details are mentioned in Refs.\cite{Emtsev2009,Kruskopf2016}.

Prior to intercalation, the samples were degassed at a temperature of 550~\degC\; until a base pressure below $5\times10^{-10}$~mbar was reached. Pb (Sigma-Aldrich, 99.999\%) was evaporated from a Knudsen cell onto the ZLG at room temperature (RT) under ultra-high vacuum (UHV) conditions, while adjusting the deposition rate with a quartz microbalance to 1--2 monolayers (ML) per minute. Intercalation was achieved by repeated cycles of deposition and annealing. The temperature during the direct current heating was monitored by an infrared pyrometer with an emissivity of 0.86. Three typical annealing steps were used: (1) \textit{mild annealing:} 350~\degC\; for a long time (several hours); (2) \textit{medium temperatures:} 500~\degC\; for 5~min; (3) \textit{high temperature:} 700~\degC\; for 5~min (occasionally flash annealing at 1000~\degC\; for 15~s).

The intercalation progress as well as the interface structures were monitored and analyzed in detail by in-situ high-resolution electron diffraction experiments using  SPA-LEED. The diffraction patterns in Fig.~\ref{FIG1}(a) and Fig.~\ref{FIG11}(a) were corrected using the LEEDCal 2013 v4.1 software \cite{Sojka2013}. Ultrahigh vacuum (UHV) SEM was used to monitor the sample surface morphology and the distribution of intercalated phases from millimeter to nanometer scale. Nanoscopic and atomic scale investigations were performed by STM. UHV conditions were preserved during the transfer of the sample between chambers.

In addition to these in-house experiments, LEEM measurements were performed at the aberration-corrected spectroscopic photoemission and low energy electron microscope station (MAXPEEM) at MAX IV in Lund, Sweden. For more details please refer to Ref.~\cite{Niu2023}. For these measurements, the sample was remounted in an Ar glovebox and degassed at 200~\degC\; for three hours. Measurements were performed at RT unless stated otherwise.

\vspace{2em}
\textbf{Acknowledgement}
We gratefully acknowledge financial support from the DFG through FOR5242 (TE386/22-1), We thank the AG Seyller (TU Chemnitz) and Ulrich Starke (MPI Stuttgart) for providing ZLG samples. 
We also acknowledge the fruitful discussions with  Ulrich Starke and Kathrin Küster (MPI Stuttgart).

\providecommand{\latin}[1]{#1}
\makeatletter
\providecommand{\doi}
  {\begingroup\let\do\@makeother\dospecials
  \catcode`\{=1 \catcode`\}=2 \doi@aux}
\providecommand{\doi@aux}[1]{\endgroup\texttt{#1}}
\makeatother
\providecommand*\mcitethebibliography{\thebibliography}
\csname @ifundefined\endcsname{endmcitethebibliography}
  {\let\endmcitethebibliography\endthebibliography}{}

\end{document}